\documentclass[journal=jacsat,manuscript=article]{achemso}

\usepackage[version=3]{mhchem} 

\usepackage{balance}
\usepackage{mathptmx}
\usepackage{amsmath}
\usepackage{sectsty}
\usepackage{graphicx} 
\usepackage{lastpage}
\usepackage{dcolumn}
\usepackage{bm}
\usepackage{float}
\usepackage{fancyhdr}
\usepackage{fnpos}
\author{Jincheng Yue}
\affiliation{Institute of High Pressure Physics, School of Physical Science and Technology, Ningbo University, Ningbo, 315211, China}
\alsoaffiliation{These authors contributed equally}

\author{Junda Li}
\affiliation{Institute of High Pressure Physics, School of Physical Science and Technology, Ningbo University, Ningbo, 315211, China}
\alsoaffiliation{These authors contributed equally}

\author{Jiongzhi Zheng}
\affiliation{Thayer School of Engineering, Dartmouth College, Hanover, New Hampshire, 03755, USA}
\alsoaffiliation[Second University]
{Department of Mechanical and Aerospace Engineering, The Hong Kong University of Science and Technology, Clear Water Bay, Kowloon, 999077, Hong Kong}
\alsoaffiliation{These authors contributed equally}
\email{jiongzhi.zheng@dartmouth.edu}

\author{Xingchen Shen}
\affiliation{Laboratoire de Cristallographie et Sciences des Matériaux (CRISMAT), CNRS, ENSICAEN, Caen, 14000, France}

\author{Wenling Ren}
\affiliation{Institute of Materials Science, Technical University of Darmstadt, Darmstadt, 64287, Germany}

\author{Yanhui Liu}
\affiliation
{Institute of High Pressure Physics, School of Physical Science and Technology, Ningbo University, Ningbo, 315211, China}
\email{liuyanhui@nbu.edu.cn}

\author{Tian Cui}
\affiliation
{Institute of High Pressure Physics, School of Physical Science and Technology, Ningbo University, Ningbo,  315211, China}
\alsoaffiliation[Second University]
{State Key Laboratory of Superhard Materials, College of Physics, Jilin University, Changchun, 130012, China}
\email{cuitian@nbu.edu.cn}

\title[An \textsf{achemso} demo]
  {Hierarchical Characterization of Thermoelectric Performance in Copper-Based Chalcogenide CsCu$_3$S$_2$: Unveiling the role of Anharmonic Lattice Dynamics}

\abbreviations{IR,NMR,UV}

\begin{document}
\begin{sloppypar}

\clearpage
\begin{abstract}
 Fundamental understanding of anharmonic lattice dynamics and heat conductance physics in crystalline compounds is critical for the development of thermoelectric energy conversion devices. Herein, we thoroughly investigate the microscopic mechanisms of thermal transport in CsCu$_3$S$_2$ by coupling the self-consistent phonon (SCP) theory with the linearized Wigner transport equation (LWTE). We explicitly consider both phonon energy shifts and broadening arising from both cubic and quartic anharmonicities, as well as diagonal/non-diagonal terms of heat flux operators in thermal conductivity. Our findings show that the strong anharmonicity of CsCu$_3$S$_2$ primarily arises from the presence of $p$-$d$ anti-bonding hybridization between Cu and S atoms, coupled with the random oscillations of Cs atoms. Notably, the competition between phonon hardening described by the loop diagram and softening induced by the bubble diagram significantly influences particle-like propagation, predominantly reflected in group velocity and energy-conservation rule. Additionally, the electrical transport properties are determined by employing the precise momentum relaxation-time approximation (MRTA). 
At high temperatures, the thermoelectric performance of $p$-type CsCu$_3$S$_2$ reaches its optimum theoretical value of 0.94 along the in-plane direction based on advanced phonon renormalization theory. In striking contrast, the harmonic approximation theory significantly overestimates the thermoelectric efficiency at the same temperatures, rendering it an impractical expectation. Conversely, the first-order renormalization approach leads to a serious underestimation of the thermoelectric properties due to the over-correction of phonon energy. Our study not only reveals the pivotal role of anharmonic lattice dynamics in accurately assessing thermoelectric properties but also underscores the potential thermoelectric applications for novel copper-based chalcogenides. 
\end{abstract}

\clearpage
\section{Introduction}
Thermoelectric materials with low costs and superior performance are currently undergoing extensive research due to their potential for widespread application in waste heat recovery across manufacturing, processing, and automotive sectors.~\cite{yin2023low, liu2023lattice, kimber2023dynamic} The figure of merit $ZT$ is commonly used to assess the conversion efficiency of thermoelectric materials,  $ZT$=$S^2$$\sigma$/$\kappa$, where $S$ and $\sigma$ represents the Seebeck coefficient and conductivity respectively; $\kappa$ consists of electronic thermal conductivity ($\kappa_e$) and lattice thermal conductivity ($\kappa_\mathrm{L}$).~\cite{dong2023challenges, he2016ultralow, yue2023pressure} Given the challenges of decoupling electron transport parameters, achieving intrinsic ultra-low $\kappa_\mathrm{L}$ has emerged as a prerequisite for attaining high thermoelectric properties.~\cite{yue2023significantly, hong2023realizing, yue2024role} Within the framework of theoretical calculations, numerous promising thermoelectric materials have been reported in recent years owing to their exceptionally low $\kappa_\mathrm{L}$ at room temperature; e.g., chalcogenide AgIn$_5$S$_8$ (0.97 W/mK)~\cite{juneja2019rattling}, BaTiS$_3$ (0.25 W/mK)~\cite{paudel2020evaluating} Cu$_2$AgBiPbS$_4$ (0.3 W/mK)~\cite{isaacs2020inverse}, oxide LaZnOPn (Pn = P, As) (2 W/mK)~\cite{einhorn2020computational}, BaBi$_2$O$_6$ (3.6 W/mK)~\cite{spooner2021babi2o6}, layered intermetallic Na$_2$MgSn (0.81 W/mK)~\cite{wang2019low},  full-Heusler Li$_2$TlBi (2.36 W/mK)~\cite{he2019designing}  and alkali-based telluride BaIn$_2$Te$_4$ (1 W/mK)~\cite{gurel2022comprehensive}. In essence, the ultra-low $\kappa_\mathrm{L}$ often correlates with strong lattice anharmonicity, which is commonly induced by distinctive electron/atomic configurations, including the lone electron pairs~\cite{wang2016high}, resonant bonding~\cite{delaire2011giant}, host-guest structures~\cite{tadano2018quartic, tadano2015impact}, and so on.

\par Recent advancements in thermoelectric materials underscore the potential of copper-based chalcogenides as highly promising candidates for thermoelectric applications.~\cite{ma2020alpha, yue2024ultralow} Compared to conventional thermoelectric materials, copper-based chalcogenides offer numerous advantages, including inherent non-toxic properties, cost-effectiveness, and environmental friendliness. For example, superionic conductor (SIC), e.g., Cu$_2$Se$_{1-x}$S$_x$~\cite{zhang2020cu}, exhibits ultra-low $\kappa_\mathrm{L}$ of $\textless$ 1 Wm$^{-1}$K$^{-1}$ attributed to that the disordered arrangement of atoms induced by the liquid-like sublattice enhances the scattering rates of main heat carriers. However, the liquid-like migration behavior tends to degrade the performance of thermoelectric devices and shorten their lifetimes. Importantly, the introduction of ionic alkali metals has been proven to limit the occurrence of the superionic phase.~\cite{ma2020alpha, ma2019cscu5s3}. Previously, Hirohiko et al.~\cite{sato1992electronic} successfully synthesized ternary copper-based alkali metal chalcogenides CsCu$_3$S$_2$. Their experimental results confirmed the thermodynamic stability of CsCu$_3$S$_2$ and revealed a significant decrease in resistivity after soaking in an ammonium solution. Additionally, Li et al. proposed a valid empirical descriptor based on high-throughput screening to predict a series of crystals with ultra-low $\kappa_\mathrm{L}$ ($\textless$ 2 Wm$^{-1}$K$^{-1}$), including the CsCu$_3$S$_2$.~\cite{li2022high} Nevertheless, the predictive models based on the harmonic approximation (HA) treatment faces challenges in predicting thermal conductivity when applied to highly anharmonic crystals because they remarkably underestimated the phonon energy.~\cite{zheng2022anharmonicity, van2016high} Meanwhile, the temperature effect on particle-like propagation properties is solely reflected in the Bose-Einstein distribution of phonons.~\cite{tadano2015self}

\par In highly anharmonic crystalline compounds, the thermal or quantum fluctuations of atoms are too large for quasiharmonic phonon theory to accurately evaluate lattice dynamics.~\cite{tadano2022first} Concurrently, the lattice dynamics is pivotal in defining key thermal transport properties, such as phonon velocities and rates of anharmonic scattering.~\cite{zheng2024unravelling, zheng2022effects} These factors are crucial in determining thermal conductivity and its dependency on temperature, as they govern the patterns of phonon transport and interaction within the crystal framework.~\cite{xia2020high, xia2020microscopic}  With the development of the phonon self-consistency theory based on the quasiparticle (QP) approximation, the effective one-body Hamiltonian of interacting phonons has been achieved to capture the accurate modeling of anharmonic lattice dynamics at finite temperatures.~\cite{tadano2022first, ravichandran2018unified} 
However, explicitly considering both cubic and quartic anharmonicities in phonon energy shifts and phonon scattering rates is still lacking in the study of thermal transport and figure of merit in copper-based chalcogenides, as well as in most other thermoelectric materials.~\cite{xie2023microscopic, xia2020particlelike} Furthermore, a unified theory of thermal transport proposed by Simoncelli et al. enables the integration of the off-diagonal elements of heat-flux operators, enriching the computational framework including accurately depicting wavelike heat conduction.~\cite{simoncelli2019unified, simoncelli2022wigner} A comprehensive investigation aimed at accurately describing thermal transport and figure of merit is not only fundamentally interesting but also paves the way to better understand and engineer the thermoelectric performance of known compounds.

\par In this work, we have systematically investigated the effect of the anharmonic lattice dynamics on the thermoelectric properties of copper-based chalcogenides CsCu$_3$S$_2$ at the atomic level. Considering that reliable modeling of lattice dynamics requires accurate treatment of lattice anharmonicity, our calculation involves second-order anharmonic self-energy Feynman diagrams. Significantly, the net frequency shift of phonon frequencies induced by third- and fourth-order anharmonicity exerts a decisive influence on thermoelectric properties. Using the unified heat transport equation which includes particle-like propagation and wave-like tunneling channels, the thermoelectric properties of $p$-type CsCu$_3$S$_2$ are estimated accurately, and the optimal theoretical value of 0.94 is obtained along the in-plane direction at 900 K. By contrast, the predicted thermoelectric value using the traditional harmonic approximation treatment is overestimated by 88\%, reaching an unrealistic expected value of 1.77. 
Conversely, only considering the first-order renormalization contribution approach underestimates the thermoelectric properties by about 15\% because of the over-correction of phonon energy. Our research highlights the pivotal influence of anharmonic lattice dynamics on accurately assessing thermoelectric properties, underscoring the necessity to understand intricate lattice dynamics for effective thermoelectric energy conversion.

\section{Methodology}
\par All the density functional theory (DFT) calculations were performed using the projected Enhanced wave (PAW) method and implemented in the Vienna Ab initio Simulation Package.~\cite{blochl1994projector, kresse1994theory, kresse1999ultrasoft} 
The exchange-correlation function adopted the Perdew-Burke-Ernzerhof (PBE) function within the generalized gradient approximation; meanwhile, the optB86b-vdW was employed to accurately describe the van der Waals (vdW) interactions.~\cite{klimevs2011van, perdew1996generalized} 
A kinetic energy cutt-off of 600 eV, coupled with a 20×20×20 Monkhorst-Pack $k$-point mesh, was adopted to ensure enough sampling of the Brillouin zone for the primitive cell. The energy and force-convergence criteria of 10$^{-8}$ eV and 10$^{-6}$ eV·Å$^{-1}$ were used for both the structural optimization and statical self-consistent DFT calculation, respectively.

\par  In terms of assessing the zero-K harmonic force constants (IFCs), the finite-displacement method was meticulously applied.~\cite{esfarjani2008method} The static DFT calculations were conducted on a robust 3×3×2 supercell containing 108 atoms and using a 4×4×4 $k$-point mesh. To train the higher-order anharmonic IFCs, a comprehensive dataset comprising 200 atomic configurations was employed. A displacement magnitude of 0.15 Å was systematically applied to each atom in randomly selected vectors to diminish structural cross-correlations, facilitated by a random-seed algorithm.~\cite{zhou2019compressive, zhou2019compressive2} The compressive sensing lattice dynamics (CSLD) framework, alongside the discerning least absolute shrinkage and selection operator (LASSO) methodology, was harnessed to filter and affirm the physically important terms of IFCs, as integrated within the ALAMODE package.~\cite{tadano2018first}

\par After obtaining the harmonic and anharmonic IFCs, the anharmonic phonon energy renormalization was carried out by applying the self-consistent phonon theory in the reciprocal space.~\cite{errea2011anharmonic} Assuming only the first-order perturbation due to the quartic anharmonicity, namely the loop diagram, the first-order self-consistent phonon (SC1) equation can be derived in the diagonal form as follows:~\cite{tadano2015self}
\begin{equation}
    \begin{aligned}
\Omega ^{2}_{q}=\omega ^{2}_{q}+2\Omega_{q}I_{q} 
    \end{aligned}
\end{equation}
\begin{equation}
    \begin{aligned}
I_{q} = \frac{1}{8N}\sum_{q'}\frac{\hbar \Phi(q;-q;q';-q')}{4 \Omega_{q} \Omega_{q'}}[1 + 2n(\Omega_{q'})]
    \end{aligned}
\end{equation}
where $\omega^{2}_{q}$ is the bare harmonic phonon frequency associated with the phonon mode $q$, and $\Omega^{2}_{q}$ is the anharmonically renormalized phonon frequency at finite temperatures. The $n$ and $\Phi(q;-q;q';-q')$ represent the Bose-Einstein distribution ($n(\omega) = 1/[exp(\hbar \omega/k_BT)-1]$), and the reciprocal representation of 4th-order IFCs, respectively. Within the framework of QP approximation and utilizing phonon energies renormalized by loop diagram, the additional negative frequency shifts ascribing from the bubble self-energy can be estimated by the following self-consistent equation:.~\cite{tadano2022first}
\begin{equation}
    \begin{aligned}
(\Omega ^{B}_{q})^2=\Omega ^{2}_{q}-2\Omega_{q}Re {\textstyle \sum_{B}^{q}} [G,\Phi _3](\Omega = \Omega ^{B}_{q})
    \end{aligned}
\end{equation}
where $\textstyle \sum_{B}^{q}[G,\Phi _3]$, $B$, and $\Phi_3$ denote the frequency-dependent bubble self-energy, 
bubble diagram, and third-order force constant, respectively. Noteworthily, the nonlinear treatment option (QPNL) was chosen to solve Eq. (3) and obtain the fully renormalized phonon energies that account for cubic and quartic anharmonicities.
\par Within the linearized Wigner transport equation (LWTE) framework, the thermal conductivity $\kappa^\mathrm{P/C}_\mathrm{L}$ can be described as:~\cite{simoncelli2019unified}
\begin{equation}
    \begin{aligned}
\kappa^\mathrm{P/C}_\mathrm{L}=&\frac{\hbar ^2}{k_BT^2VN_0} \sum_{q}\sum_{j,j'}\frac{\Omega _{qj}+\Omega _{qj'}}{2} \upsilon _{qjj'} \otimes \upsilon _{qjj'} \\ 
&\times \frac{\Omega _{qj}n_{qj}(n_{qj}+1)+\Omega _{qj'}n_{qj'}(n_{qj'}+1)}{4(\Omega _{qj}-\Omega _{qj'})^2+ (\Gamma_{qj}+\Gamma_{qj'})^2}(\Gamma_{qj}+\Gamma_{qj'}) 
    \end{aligned}
\end{equation}
where the $V$ and $N_0$ denote the primitive-cell volume and number of sampled, respectively. The $\Gamma_{q}$ stands for the scattering rates including three-phonon (3ph), four-phonon (4ph), and isotope-phonon scattering processes. The phonon group velocity $\upsilon _{q}$ is extended to encompass off-diagonal components:~\cite{simoncelli2022wigner}
\begin{equation}
    \begin{aligned}
\upsilon _{qj'j}=\frac{\left \langle e_{qj}\left | \frac{ \partial D(q)}{\partial q}  \right | e_{qj'} \right \rangle }{2\sqrt{\Omega _{qj}\Omega _{qj'}} } 
    \end{aligned}
\end{equation}
in which the $D(q)$ and $e_{q}$ are the dynamical matrix and polarization vector, respectively. The particle-like phonon thermal transport properties are computed by using the ShengBTE and FourPhonon packages, while the wave-like tunneling of phonon channels is evaluated using our in-house code.~\cite{li2014shengbte, han2022fourphonon}
\par The related electronic transport was performed using the AMSET code, wherein the differential scattering rate from the initial $\psi_{\textit{nk}}$ to final states $\psi_{\textit{mk+q}}$ for inelastic and elastic processes was calculated as:~\cite{ganose2021efficient}
\begin{equation}   
\begin{aligned}
\tau^{-1}_{nk \rightarrow mk+q}=&\frac{2\pi}{\hbar}|g_{nm}(k,q)|^2\\
&\times[(n_{po}+1-f_{mk+q})\delta (\varepsilon _{nk}-\varepsilon _{mk+q}-\hbar \omega_{po})\\
&+(n_{po}+f_{mk+q})\delta (\varepsilon _{nk}-\varepsilon _{mk+q}+\hbar \omega_{po})]
\end{aligned}
\end{equation}
\begin{equation} 
\tilde{\tau}^{-1}_{\textit{nk} \rightarrow \textit{mk+q}} = \frac{2\pi}{\hbar}|g_{nm}(k,q)|^2\delta(\varepsilon_{nk}-\varepsilon_{mk+q})
\end{equation}
where $\varepsilon_{nk}$ symbolizes the specific energy state $\psi_{\textit{nk}}$. The $g_{nm}(k,q)$ accounts for three kinds of electron-phonon scattering matrix elements including acoustic deformation potential (ADP), polar optical phonon (POP), and ionized impurity (IMP) matrix element. Notably, the overall elastic rates were calculated using the momentum relaxation time approximation (MRTA), given by:~\cite{claes2022assessing}
\begin{equation} 
\tilde{\tau}^{-1}_{\textit{nk}} = \sum_{m}\int \frac{d}{\Omega_{BZ}}[1-\frac{\upsilon_{nk}\cdot\upsilon_{mk+q}}{|\upsilon_{nk}|^2}\tilde{\tau}^{-1}_{\textit{nk} \rightarrow \textit{mk+q}} ]
\end{equation}
\section{Results and Discussion}
\par The CsCu$_3$S$_2$ crystallizes in the trigonal space group $P\overline{3}m1$ (No. 164), in which the corresponding Wyckoff positions of Cs, Cu, and S atoms are 1b, 3e, and 2d sites, respectively. The configurations of CsCu$_3$S$_2$ are originated from the [Cu$_3$S$_2$]$^-$ pyramidal-like structural motif and independent Cs$^+$ ions. These [Cu$_3$S$_2$]$^-$ units pass through two opposite sides with two-coordinated Cu atoms to form a wave layer and propagate periodically along with the in-plane direction; meanwhile, they are interlaced with Cs$^+$ ion layers and piled up along with the out-of-plane direction, as shown in Figures 1(a-b). According to the charge density equipotential surface, the overlapping charge density is mainly distributed between Cu and S atoms, while the valence electron cloud of the Cs$^+$ ions is confined to its nucleus surface. Additionally, the predominant distribution of overlapping charge densities is observed within the layers of [Cu$_3$S$_2$]$^-$, indicating that the bonding character between Cu and S atoms exhibits a pronounced inclination towards covalent interactions. In contrast, the valence electrons of the Cs$^+$ ions are predominantly localized at the periphery of the atomic nucleus, suggesting that the electrostatic interaction between Cs$^+$ ions and [Cu$_3$S$_2$]$^-$ structures is relatively weak.

\par Uncovering the weak interatomic interaction can be done intuitively through noncovalent (NCL) interaction analysis.~\cite{johnson2010revealing} The NCI index can be effectively quantified using the Interaction Region Indicator (IRI), which operates as a function characterized by sign($\lambda_2$)$\rho$.~\cite{lu2021interaction} In this expression, $\rho$ denotes the electron density, while sign($\lambda_2$) reflects the sign associated with the second eigenvalue in the electron density Hessian matrix. Within regions characterized by diminished electron density ( $\left | \mathrm{sign} (\lambda_2)\rho \right |  < $ 0.01 a.u.), the weak interaction is typically delineated by salient peaks, which signal significant alterations in the relative differential gradient (RDG) as it approaches the zero threshold at the critical point, as demonstrated by Figure 1(c). Simultaneously, the three-dimensional IRI isosurfaces, highlighted by BGR color bars, demonstrate that the weak vdW forces are primarily localized around the Cs$^+$ ions. As a result, the Cs$^+$ ions display stochastic oscillation behavior similar to rattles, accompanied by larger mean squared atomic displacements (MSDs). [see Figure S1] In Figure 1(d), we conducted a detailed examination of the orbital-resolved partial density of states (PDOS) and crystal orbital Hamiltonian populations (COHP) to elucidate the characteristics of the Cu-S bonding interactions. The Cu($d$)-orbitals converge energetically with the S($p$)-orbitals, fostering a robust $p$-$d$ hybridization with a wide energy range.~\cite{jaffe1984theory, jaffe1983anion} Such strong hybridization propels the occupied antibonding states up to the Fermi level, diminishing the bond strength and manifesting metavalent-like bonding characteristics.~\cite{gholami2024unlocking} On the whole, the hybridization of $p$-$d$ antibonding orbitals between Cu and S atoms, in conjunction with the stochastic vibrations of Cs$^+$ ions, gives rise to potentially strong lattice anharmonicity.~\cite{yue2024ultralow}

\par Figure 2(a) shows the anharmonically renormalized phonon dispersions at a finite temperature in comparison with those obtained from harmonic approximation. Obviously, the acoustic and mid-frequency optical branches ($<$ 180 cm$^{-1}$) undergo a notable hardening, whereas the high-frequency optical branches exhibit a non-negligible softening as temperature increases. It is essential to note the significance of both cubic and quartic interatomic force constants in accurately predicting the lattice dynamics within materials exhibiting strong anharmonicity. In general, the frequency shift caused by anharmonicity at a finite temperature is represented by the real part of the phonon self-energy.~\cite{tadano2015impact, tadano2015self} Specifically, considering the first-order corrections arising from fourth-order anharmonicity (SCPH), the SCPH method typically generates a positive contribution to the phonon frequency. Nevertheless, the SCPH theory tends to overestimate phonon frequencies, a tendency predominantly because of the neglection of the negative frequency shift associated with bubble self-energy.~\cite{tadano2022first, zheng2022anharmonicity} Subsequently, we compare the temperature-dependent phonon frequency difference of TO$_{5th}$ mode at $\Gamma$ point, including the loop diagram with the polarization mixing (PM), and that of the bubble diagram respectively. [see Figure S2] Indeed, the observed negative shift is considerable in the highly anharmonic CsCu$_3$S$_2$, which exerts an essential effect on the theoretical prediction of phonon linewidth and consequent thermoelectric properties, as we discuss later.

\par Besides, we further delve deeper to uncover the fundamental characteristics of its lattice dynamics. Along the G-M direction, an evident avoided-crossing interaction is discernible between the longitudinal acoustic (LA) branches and low-lying optical (LLO) branches, which leads to a substantial influence on the heat-carrying phonon transport within this region.~\cite{christensen2008avoided} To begin with, the presence of the avoided-crossing point is expected to suppress the group velocity around the LA branch, akin to observations from the inelastic neutron scattering analysis of Ba$_8$Ga$_{16}$Ge$_{30}$.~\cite{xie2023microscopic, tadano2018first} The LA branch, which is initially the most heavily colored, is noticeably lightened after passing through the crossing point. [see Figure S3] Additionally, there is a marked hybridization between the optical and acoustic vibrational states near the avoided-crossing point, meaning strong coupling of these phonon modes. By hybridizing the optical phonon eigenvectors into acoustic phonons, the scattering rates of the LA branch increase by several orders of magnitude, even at low-frequency limits.~\cite{li2016influence} On the other hand, the atomic participation ratio (APR) serves as a quantitative degree to assess the involvement of atoms in a specific phonon mode,
\begin{equation} 
APR_{qj,i}=\frac{|e(i;qj)|^2}{M_i}/(N\sum_{i=1}^{N} \frac{|e(i;qj)|^4}{M_i^2}) ^{1/2}
\end{equation}
where the $e$, $i$, $M$, and $N$ represent phonon eigenvector, atomic index, atomic mass, and total number of atoms, respectively. In Figures 2(b-e), it is observed that the low-energy Einstein optical modes participating in the eigenvector hybridization, predominantly originate from the oscillations of Cs$^+$ ions. Numerous flattened low-energy optical branches remarkably assist in the scattering of acoustic phonons responsible for heat transport.~\cite{li2023wavelike} By contrast, the Cu atoms predominantly contribute to the acoustic phonons and mid-frequency optical phonons, whereas the S atom exerts a dominant influence over the high-frequency optical phonon modes.
\par Afterwards, the investigation delves into the influence of anharmonic lattice dynamics on the thermal conductivity $\kappa_\mathrm{L}$ within the CsCu$_3$S$_2$ crystal, as presented in Figure 3(a). For particle-like propagation, the lattice thermal conductivities $\kappa_\mathrm{L}^\mathrm{P}$ obtained from the HA+3,4ph model, are 1.24 and 0.88 Wm$^{-1}$K$^{-1}$, respectively. Remarkably, the thermal energy is constrained within the [Cu$_3$S$_2$]$^-$ layers owing to the "quasi-stationary" nature of Cs$^+$ ions [see Figure 2(b)], facilitating its directional transport primarily along the in-plane direction ($x$-axis).~\cite{zeng2022extreme} After considering the first-order correction of the fourth-order anharmonicity to the dynamic matrix, namely, the SCPH+3,4ph model, the predicted $\kappa_\mathrm{L}^\mathrm{P}$ reaches 2.20 and 1.65 Wm$^{-1}$K$^{-1}$. We attribute this to the marked enhancement in the phonon group velocities coupled with a prolonged phonon relaxation time. [see Figure 3(b)] Incorporating the effects of third-order anharmonicity, namely, the SCPB+3,4ph model, the negative frequency shift further leads to the decrease of thermal conductivity, with the values descending to 1.80 and 1.37 Wm$^{-1}$K$^{-1}$, respectively. 
Correspondingly, both the group velocity and phonon relaxation time exhibit an analogous trend in variation, underscoring the consistent nature of their respective evolutions within the framework of the dynamics.  Significantly, the determinants of phonon lifetime under diverse theories are exclusively determined by the strict energy conservation laws.~\cite{ravichandran2020phonon} As illustrated in Figures 3(c-d), the Peierls's thermal conductivity $\kappa_\mathrm{L}^\mathrm{P}$ is almost entirely contributed by phonons below 100 cm$^{-1}$. Meanwhile, it also reveals that the competitive relationship between phonon hardening described by the loop diagram and softening induced by the bubble diagram has a decisive influence on particle propagation.~\cite{zheng2022anharmonicity}
\par Temperature-dependent fluctuations in thermal conductivity $\kappa_\mathrm{L}^\mathrm{P}$ can be attributed to anharmonic lattice dynamics as well. As anticipated, considering the 3,4ph scattering process in conjunction with the zero-K second-order force constant, derived from the HA+3,4ph model, the $\kappa_\mathrm{L}^\mathrm{P}$ follows the temperature dependence of $T^{-1.37/-1.34}$. Such a trend diverges from the conventional $T^{-1}$ relationship as the introduction of additional 4ph scattering tends to enhance the temperature-dependence of $\kappa_\mathrm{L}^\mathrm{P}$.~\cite{zhu2020violation, xia2020microscopic} [see Figure S4] The SCPH+3,4ph method using first-order renormalization drastically weakens the temperature dependence of thermal conductivity to $T^{-0.86/-0.78}$. In contrast to the SCPH+3,4ph model, the SCPB+3,4ph model gives rise to a faster decay of $T^{-1.05/-1.07}$ via further including the negative frequency shift. It is essential to underscore that while the temperature dependence exhibited under the SCPB+3,4ph model aligns with the outcomes of the lowest-order theoretical HA+3ph model, the anharmonic lattice dynamics still have an irreplaceable role in accurately predicting the thermal conductivity.

\par Traditionally, in Peierls-Boltzmann transport theory, coherence terms—specifically, the off-diagonal elements of heat flux operators—are disregarded, a valid assumption when phonon interbranch spacings significantly exceed linewidths.~\cite{simoncelli2019unified, simoncelli2022wigner}  Nonetheless, as the phonon mean free path (MFP) nears the interatomic spacing, these off-diagonal contributions become non-negligible and amenable to calculation through a unified framework of thermal transport.~\cite{zheng2022anharmonicity} [see Figure 3(a) and Eq.(4)] Generally, phonons with a lifetime below the Wigner time limit $[\Delta \omega_{av}]^{-1}$ (red solid line in Figure 3(b)), mainly contribute to coherence conductivity $\kappa_\mathrm{L}^\mathrm{C}$ via wave-like tunneling. [see Figure 3(b)] As indicated by Figures 4(a-d), the coherence conductivity $\kappa_\mathrm{L}^\mathrm{C}$ primarily originates from the quasi-coalescing vibrations $(\omega_{qj} \cong  \omega_{qj})$ within the low- to intermediate-frequency region ($<$ 180 cm$^{-1}$). In particular, flat low-energy optical branch (50-100 cm$^{-1}$) primarily driven by the Cu$^+$ and rattling-like Cs$^+$ ions are correlated with strong 4ph scattering, which tends to inhibit particle-like thermal transport and promotes coherence among non-degenerate states.~\cite{zheng2024unravelling} [see Figure 2(c-d)] On the contrary, high-frequency optical phonons generate negligible influence on both particle-like propagation and wave-like tunneling, primarily due to their extremely flat dispersion characteristics. [see Figures 3(b-f) and Figure 2(b)]

\par The theoretically predicted ultra-low thermal conductivity $\kappa_\mathrm{L}$ of CsCu$_3$S$_2$ underscores its prospects for integration into thermoelectric devices. Consequently, we thoroughly investigate its electron structure and transport properties within the framework of the momentum relaxation time approximation (MRTA).~\cite{ganose2021efficient}
As mentioned earlier, the hybridization of Cu($d$)-S($p$) orbitals produces an antibonding state below the Fermi level, which also determines the band structure and electron transport properties.~\cite{yue2024ultralow, gholami2024unlocking} [see Figure 1(d)] Our computation reveals an indirect band gap quantified at 1.6 eV using PBE (HSE06: 2.6 eV), positioning the valence band maximum (VBM) at the A point, accompanied by a double degeneracy. [see Figure 5(a)] The integration of dual-band transmission with the spatial multi-valley convergence at point A facilitates the enhancement of electrical conductivity and the Seebeck coefficient, thereby guaranteeing that both parameters are optimized and intended for improving thermoelectric performance.~\cite{pei2011convergence, tang2015convergence} [see Figure 5(b)]

\par Moreover, the specific type of doping is a priority when thermoelectric materials are synthesized industrially, while the likelihood of $p$-type or $n$-type doping can be inferred from the ionization potentials (IP) and electron affinity (EA).~\cite{brlec20222} Generally, the lower IP implies that the low ionization potentials from high-lying valence band maxima drive semiconducting materials to produce holes.~\cite{wei1998calculated} The vacuum and core energies were used to compute the band alignment for two materials according to the core-vacuum alignment scheme.
\begin{equation}
    IP = (E_{vac}-E_{core, slab})-(E_{VBM}-E_{core, bulk})
\end{equation}
where the $E_{vac}$ presents the energies of the vacuum and the 
$E_{core, slab}$ denotes the core level in the bulk-like surface slab. The $E_{VBM}$ and $E_{core, bulk}$ are the valence band maximum and core energy of the bulk, respectively. The Cu 2$p$ state was used for aligning the energy levels, which has been widely used in previous copper-based chalcogenide (CBC) compounds, and the (001) slab was utilized to obtain the energies of the surface.~\cite{ghorbani2020efficiency} As demonstrated in Figure 5(c), the calculated IP of CsCu$_3$S$_2$ with 3.6 eV, aligns with that of conventional $p$-type semiconductors such as LaZnOAs (3.6 eV), and LaZnOP (3.9 eV).~\cite{einhorn2020computational} Accordingly, its low-lying IP is indicative of a propensity for hole doping, which demonstrates its considerable promise as a $p$-type thermoelectric material.
\par Figures 6(a-f) present a comprehensive depiction of the thermal variation in electron properties such as electrical conductivity $\sigma$, Seebeck coefficient $S$, and electronic thermal conductivity $\kappa_e$ for CsCu$_3$S$_2$ ranging from 10$^{18}$ to 10$^{21}$ cm$^{-3}$ and temperatures spanning from 300 to 900 K. The VBM is chiefly characterized by the orbital intermingling of Cu and S atoms, resulting in conductance being predominantly conveyed across the in-plane within the [Cu$_3$S$_2$]$^-$ layers, as evidenced by the carrier mobility. [see Figure S5] Within the carrier scattering mechanisms, the dominance of inelastic interactions associated with polar optical phonons constitutes a significant impediment to carrier mobility. [see Figure S6] The observed conductivity $\sigma$ exhibits an inverse relationship with temperature while maintaining a direct proportionality with the carrier concentration. Similarly, the electronic thermal conductivity $\kappa_e$ parallels the conductivity $\sigma$ in its correlation with temperature and carrier concentration, a relationship that is consistent with the principles of the Wiedemann-Franz law.~\cite{lavasani2019wiedemann} On the contrary, the Seebeck coefficient $S$ is determined by the ratio of the first moment of the generalized transport coefficient to the conductivity $\sigma$, thereby exhibiting an inverse dependency on both temperature and carrier concentration.~\cite{madsen2018boltztrap2} In situations of strong electrical transport parameter coupling, the power factor $S^2 \sigma$ is predicted to reach its maximum value of 1.25 mWm$^{-1}$K$^{-1}$ along with the in-plane direction at 900 K and the carrier concentration of 10$^{21}$ cm$^{-1}$.

\par Subsequently, our investigation explores the influences that anharmonic lattice dynamics exert on the thermoelectric properties, as exemplified by Figure 7(a-f). Under the carrier concentration of 1.49 $\times$ 10$^{20}$ cm$^{-1}$ and temperature of 900 K, the HA+3,4ph model, gives a theoretical maximum of 1.77 along with the in-plane direction. Such impressive value even qualifies this material as a high-caliber candidate in the realm of thermoelectric materials. However, the first-order SCPH+3,4ph theory indicates that the optimal value is only 0.8, corresponding to a carrier concentration of 3.04 $\times$ 10$^{20}$ cm$^{-1}$ at the same temperature. The overcorrection of the fourth-order anharmonicity on phonon energies is chiefly responsible for the virtual high of thermal conductivity $\kappa_\mathrm{L}$ and the consequent underperformance in thermoelectric properties. Further, the second-order SCPB+3,4ph model combined with the OD term further improves the optimal $ZT$ values to 0.94 with a carrier of 2.4 $\times$ 10$^{20}$ cm$^{-1}$. Emphatically, although the $ZT$ values derived from the SCPB+3,4ph+OD model may not be particularly remarkable, they are physically correct. In particular, the incorporation of sufficient anharmonic lattice dynamics is crucial for precisely determining the optimal thermoelectric properties and doping levels, thereby advancing the experimental exploration of thermoelectric materials with strong anharmonicity.
\section{Conclusions}
In conclusion, we have systematically investigated the influences of anharmonic lattice dynamics on the thermoelectric properties within the crystalline CsCu$_3$S$_2$. Initially, our findings demonstrate that the strong anharmonicity of the CsCu$_3$S$_2$ is driven by the $p$-$d$ antibonding hybridization occurring between the Cu and S atoms and the stochastic vibrations of Cs ions. Subsequently, we compare the lattice dynamics within HA, first-order SCPH, and second-order SCPB frameworks, examining the particle-like thermal conduction considering three- and four-phonon scattering. Especially, it is discovered that the anharmonic renormalization, stemming from both cubic and quartic anharmonicities, plays a vital role in the accurate depiction of thermal transport. The competition between the loop diagram and bubble diagram for phonon energy determines the particle-like group velocity and energy conservation rules, which indirectly affect the thermoelectric properties. Furthermore, the low ionization potentials within the CsCu$_3$S$_2$ render it more amenable to $p$-type doping following the establishment of band alignment. The traditional HA+3,4ph theory tends to considerably overestimate the thermoelectric efficiency, reaching 1.77 along with in-plane direction at 900 K, which results in setting impractical expectations. Conversely, the overcorrection of phonon self-energy by the one-order SCPH+3,4ph model results in a serious underestimation of the thermoelectric properties.
In contrast, by employing the advanced renormalization theory, the second-order SCPB+3,4ph in conjunction with the OD term, we can derive thermal conductivity that is physically plausible, alongside the related thermoelectric characteristics. At 900 K, the thermoelectric properties of $p$-type CsCu$_3$S$_2$ reach its optimum theoretical value of 0.94 along the in-plane direction. Our research highlights the pivotal influence of anharmonic lattice dynamics on the precise assessment of thermoelectric characteristics, providing a reliable framework for experimental investigations into the thermoelectric behavior of highly anharmonic crystals.

\section*{Author Contributions}
\par \noindent The authors confirm their contribution to the paper as follows: writing - original draft $:$ J.C.Y. and J.Z.Z.; conceptualization $:$ J.D.L; data curation $:$ J.C.Y. and W.L.R; investigation$:$ X.C.S; writing - review $\&$ editing $:$ Y.H.L and T.C.; funding acquisition $:$ T.C. All authors reviewed the results and approved the final version of the manuscript.

\section*{Conflicts of interest}
\par \noindent  The authors declare that they have no known competing financial
interests or personal relationships that could have appeared to influence the work reported in this paper

\begin{acknowledgement}
\par \noindent This work was supported by the National Natural Science Foundation of China (Grant No. 52072188, No. 12204254), the Program for Science and Technology Innovation Team in Zhejiang (Grant No. 2021R01004), 
and the Natural Science Foundation of Zhejiang province (Grant No. LQ23A040005). We are grateful to the Institute of High-pressure Physics of Ningbo University for its computational resources.
\end{acknowledgement}

\bibliography{achemso-demo}

\clearpage
\begin{figure}[htbp!] 
    \begin{center}
    \includegraphics[height=8cm]{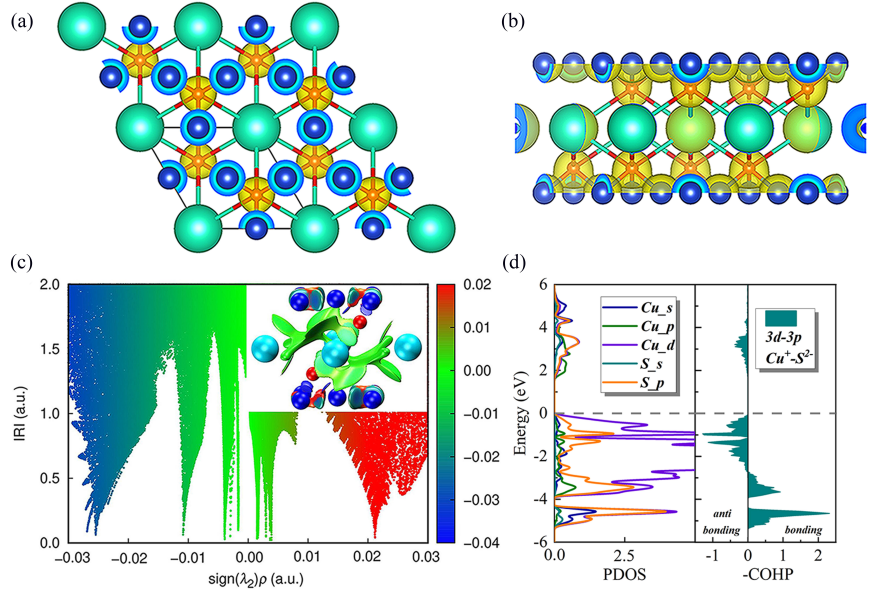}
    \caption{Structure configurations coupled with charge distribution form the (a) in-plane and (b) out-of-plane perspectives. The cyan, blue, and red spheres denote cesium, copper, and sulfur atoms, respectively. (c) Non-covalent interaction analysis associated with interaction region indicator (IRI). Inset: Isosurface map with standard coloring method and chemical explanation of sign($\lambda_2$)$\rho$ on IRI isosurfaces. (d) Orbital-resolved projected density of states (PDOS) and crystal orbital Hamilton populations (COHP) of the Cu$^+$-S$^{2-}$ orbital interaction.} 
    \end{center}
 \end{figure}
 \clearpage
 \begin{figure*}[t]
    \begin{center}
    \includegraphics[height=7cm]{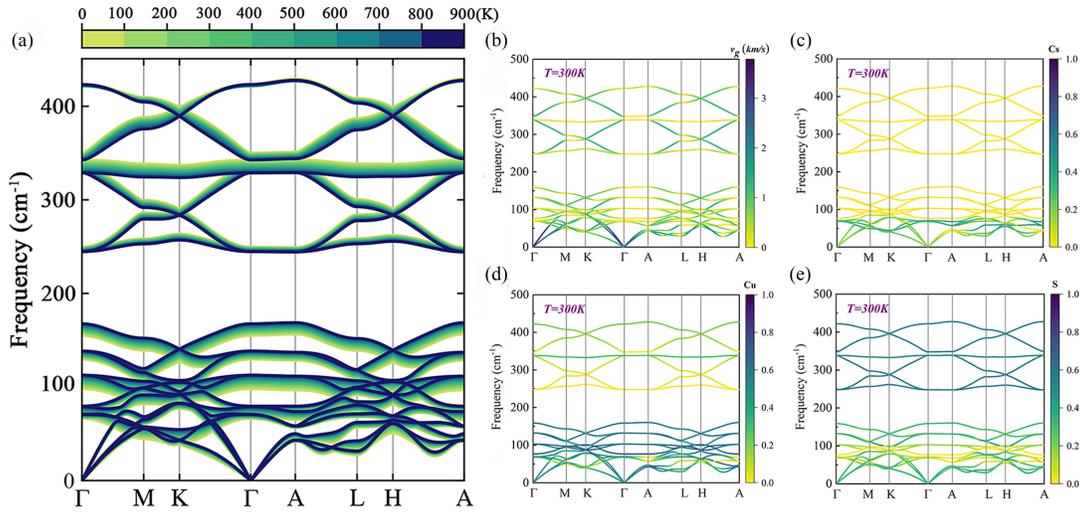}
    \caption{(a) Anharmonically renormalized phonon dispersions at finite temperatures in comparison with harmonic approximation (HA). 
    (b) Phonon group velocity projection onto the phonon dispersions at 300 K. (c) Color-coded atomic participation ratio (APR) of Cs atoms projected onto the phonon dispersions at 300 K. (d) The same as (c) but for Cu atoms. (e) The same as (c) but for S atoms.} 
    \end{center}
 \end{figure*}
 \clearpage
\begin{figure}[t]
    \begin{center}
    \includegraphics[height=14cm]{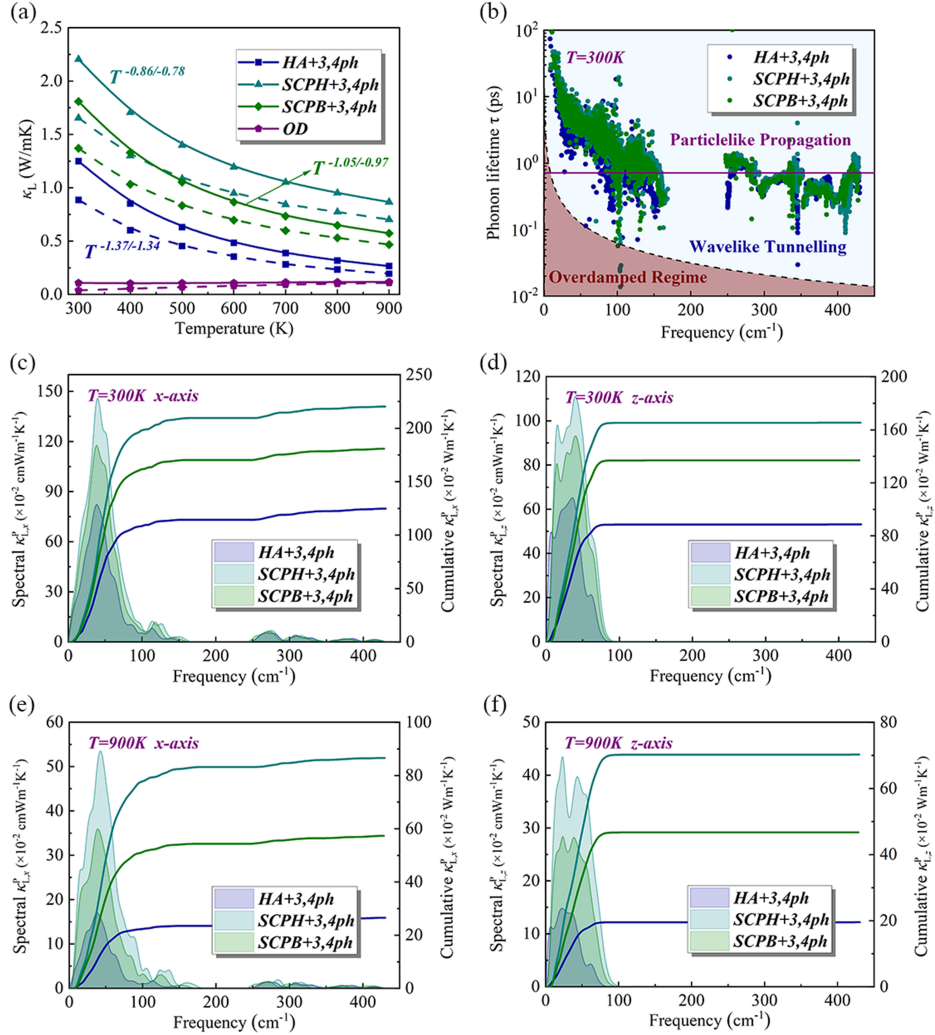}
    \caption{(a) Temperature-dependence of lattice thermal conductivity using various theories including HA+3/4ph, SCPH+3/4ph, SCPB+3/4ph, and OD model. The solid and dotted lines correspond to the $x$ and $z$ directions, respectively. (b) Calculated phonon lifetime as a function of frequency at 300 K, where the black dotted
    line and red solid line represent the Ioffe-Regel limit and Wigner limit in time. (c) Calculated spectral/cumulative populations’ thermal conductivity along the $x$-axis at 300 K. (d) The same as (c) but for $z$-axis. (e) The same as (c) but for 900 K. (f) The same as (d) but for 900 K. } 
    \end{center}
 \end{figure}
 \clearpage
 \begin{figure}[t]
    \begin{center}
    \includegraphics[height=10cm]{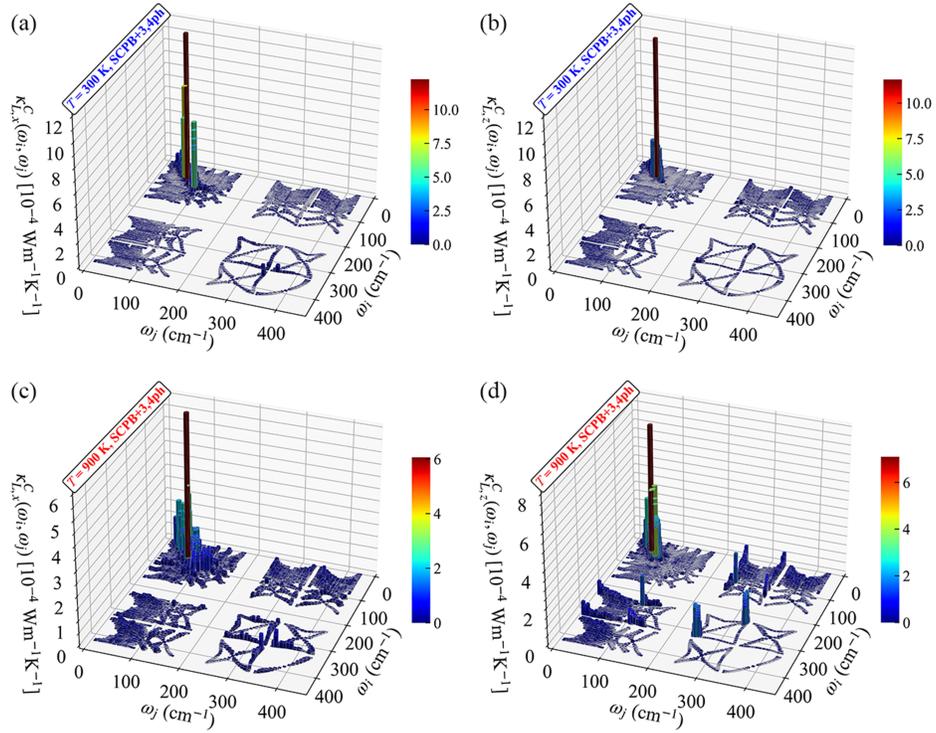}
    \caption{(a) Three-dimensional visualizations $\kappa^\mathrm{C}_\mathrm{L}$($\omega_{qj}$,$\omega_{qj'}$) of the coherences' thermal conductivity based on the SCPB+3,4ph model along with the $x$-axis at 300 K. The diagonal data points ($\omega_{qj}$ = $\omega_{qj'}$) indicate phonon degenerate eigenstates. (b) The same as (a), but for $z$-axis. (c) The same as (a), but for 800 K. (d) The same as (b), but for 800 K.} 
    \end{center}
 \end{figure}
 \clearpage
 \begin{figure}[t]
    \begin{center}
    \includegraphics[height=8cm]{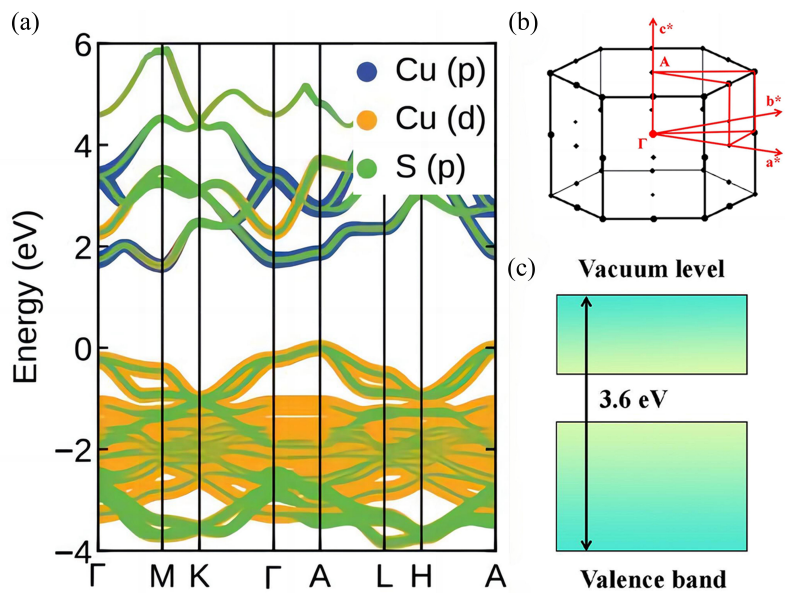}
    \caption{(a) Atom-projected electronic band structures for CsCu$_3$S$_2$. The blue, orange, and green projections represent the contributions of the Cu($p$), Cu($d$), and S($p$) orbitals, respectively. (b) First Brillouin zone and high-symmetry $k$-path of CsCu$_3$S$_2$. (c) Band alignment for CsCu$_3$S$_2$ with energies calculated relative to the Cu 2$p$ state.} 
    \end{center}
 \end{figure}
 \clearpage
  \begin{figure*}[t]
    \begin{center}
    \includegraphics[height=8cm]{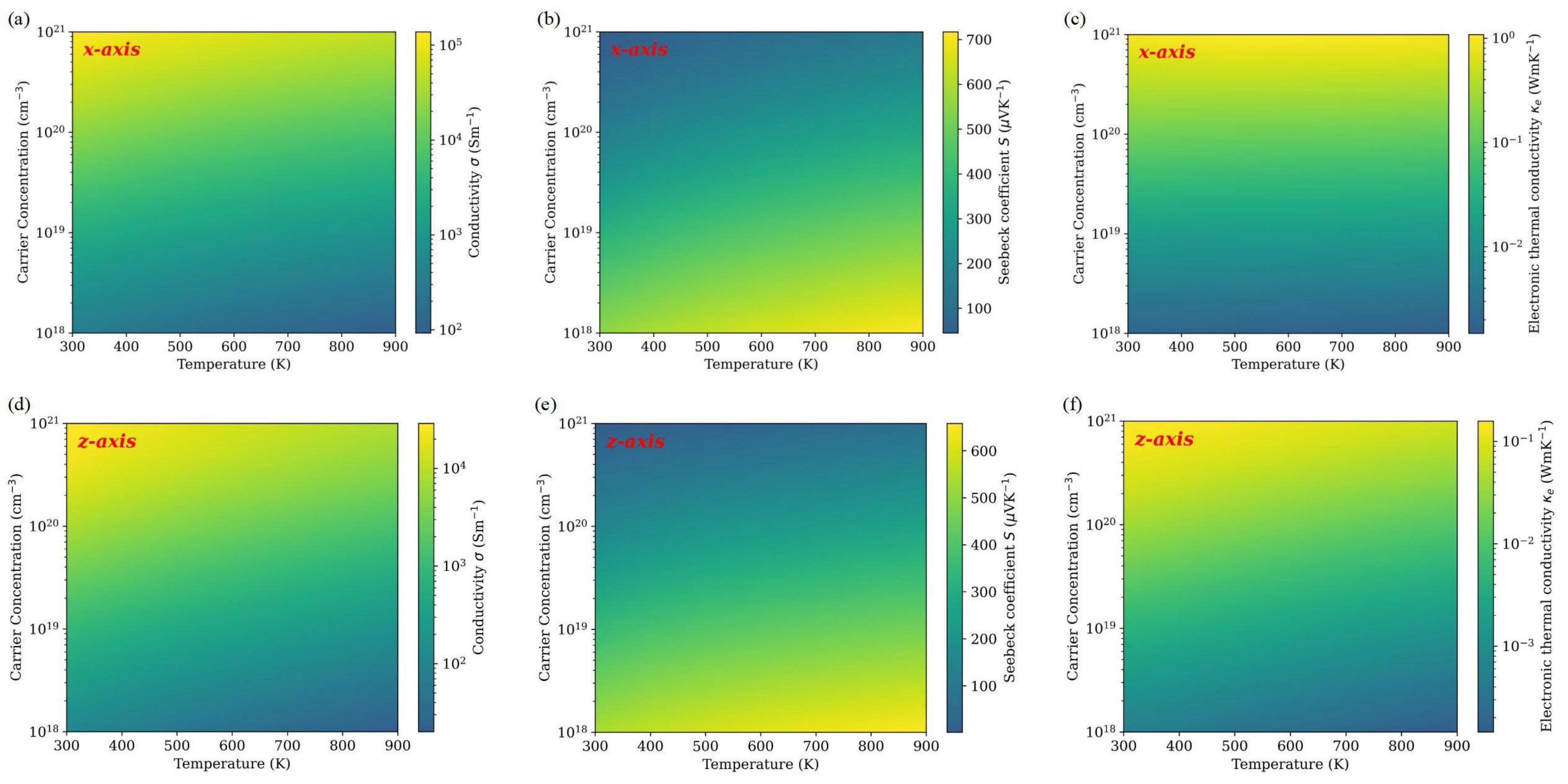}
    \caption{Calculated $p$-type electronic transport properties including the (a) conductivity $\sigma$, (b) Seebeck coefficient $S$, and (c) electrical thermal conductivity $\kappa_\mathrm{e}$ along with the $x$-axis, corresponding to carrier concentrations ranging from 10$^{18}$ to 10$^{21}$ cm$^{-3}$ and temperatures from 300 to 900 K. (d) The same as (a), but for $z$-axis. (e) The same as (b), but for $z$-axis. (f) The same as (c), but for $z$-axis. } 
    \end{center}
 \end{figure*}
 \clearpage
 \begin{figure*}[t]
    \begin{center}
    \includegraphics[height=8cm]{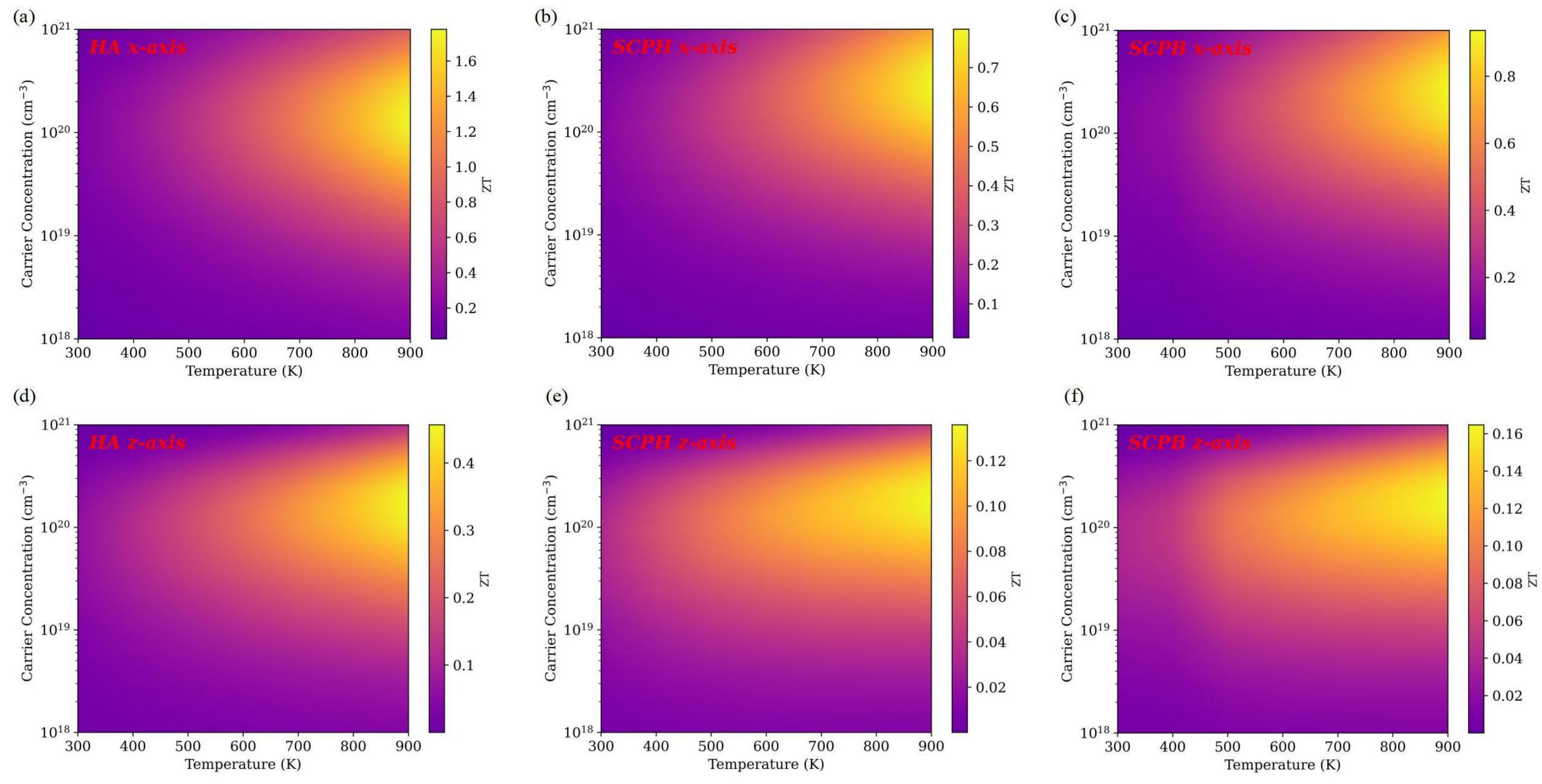}
    \caption{Calculated $p$-type thermoelectric figure of merit $ZT$ along with the $x$-axis, based on the (a) HA, (b) SCPH, and (c) SCPB+OD model, respectively, corresponding to carrier concentrations ranging from 10$^{18}$ to 10$^{21}$ cm$^{-3}$ and temperatures spanning from 300 to 900 K. (d) The same as (a), but for $z$-axis. (e) The same as (b), but for $z$-axis. (f) The same as (c), but for $z$-axis.}
    \end{center}
 \end{figure*}
 \clearpage
\end{sloppypar}
\end{document}